\documentclass[aps,nofootinbib,showpacs,noshowkeys,preprintnumbers,amsmath,amssymb]{revtex4}
\usepackage{graphicx}

\usepackage{CJK}

\usepackage{hyperref}
\hypersetup{colorlinks,linkcolor=blue,citecolor=blue,urlcolor=blue}

\allowdisplaybreaks[4]

\begin{document}


\title{Axion-Photon Conversion of LHAASO Multi-TeV and PeV Photons}\thanks{Preprint  \href{http://arxiv.org/abs/2210.13120 }{arXiv:2210.13120}, published in Chinese Phys. Lett. 40 (2023) 011401, \href{https://doi.org/10.1088/0256-307X/40/1/011401}{https://doi.org/10.1088/0256-307X/40/1/011401}}

\author{Guangshuai Zhang (\begin{CJK}{UTF8}{gbsn}张光帅\end{CJK})$^1$}

\author{Bo-Qiang Ma (\begin{CJK}{UTF8}{gbsn}马伯强\end{CJK})$^{1,2,3}$}
\thanks{Corresponding author. Email: mabq@pku.edu.cn}

\affiliation{$^1$School of Physics, 
Peking University, Beijing 100871,
	China\\
	$^2$Center for High Energy Physics, Peking University, Beijing 100871, China\\
$^3$Collaborative Innovation Center of Quantum Matter, Beijing, China}

\begin{abstract}
The Large High Altitude Air Shower Observatory (LHAASO) has reported the detection of a large number of multi-TeV-scale photon events including also several PeV-scale gamma-ray-photon events with energy as high as 1.4~PeV. The possibility that some of these events may have extragalactic origins is not yet excluded. Here we propose a mechanism for the traveling of very-high-energy (VHE) and ultra-high-energy (UHE) photons based upon the axion-photon conversion scenario, which allows extragalactic above-threshold photons to be detected by observers on the Earth. We show that the axion-photon conversation can serve as an alternative mechanism, besides the threshold anomaly due to Lorentz invariance violation, for the very-high-energy features of the newly observed gamma ray burst GRB 221009A.

\end{abstract}
\keywords{high-energy photon; axion; axion-photon conversion; gamma ray burst}

\pacs{14.80.Va; 
11.30.Cp;  
12.60.-i; 
98.70.Rz} 

\maketitle

It has long been known that the Universe is opaque to very-high-energy (VHE) photons with energy above 100~GeV and ultra-high-energy (UHE) photons with energy above 100~TeV, since these photons may collide with low-energy background photons, 
such as cosmic microwave background (CMB) photons or extragalactic background light (EBL) photons, and produce electron-positron pairs \cite{Gould:1966pza, Fazio:1970pr, Protheroe:2000hp} during their propagation in the Universe. In other words, gamma ray photons whose energy is above certain threshold cannot travel very long in the Universe due to the absorption of background low-energy photons. 

Suppose a UHE photon and a low-energy background photon collide with each other and produce the electron-positron pair. For the process to be kinematically possible, it requires that
\begin{equation}
s > 4 m_e^2 c^4,    
\end{equation}
where $s = (p_{\gamma} + p_{b})^2 = (p_{e^+} + p_{e^-})^2$ is the invariant mass of the process. It follows that, for given background photon energy, the energy of the gamma ray photon should exceed certain threshold,
\begin{equation}
    E_{\gamma} > E_{\mathrm{th}} \equiv \frac{2 m_e^2 c^4}{\epsilon (1 - \cos \theta)},
\end{equation}
where $E_{\gamma}$ is the energy of the gamma ray photon, $\epsilon$ is the energy of the soft photon and $\theta$ is the angle between the three-momenta of the two photons in the observer frame. Conversely, 
\begin{equation}
    \epsilon > \epsilon_{\mathrm{th}} \equiv \frac{2 m_e^2 c^4}{E (1 - \cos \theta)}. 
\end{equation}
 The cross section for the process $\gamma\gamma \rightarrow e^+e^-$ in tree level is well-known, 
\begin{equation}
    \sigma(v) = \frac{\pi}{2}\frac{\alpha^2}{m_e^2}(1 - v^2)\Big[(3 - v^4)\ln\frac{1+v}{1-v} - 2v(2 - v^2)\Big],
\end{equation}
which is the so-called Breit-Wheeler formula~\cite{Breit:1934zz}. Here $v$ is the speed of the outgoing electron-positron in the center of mass frame of the process
\begin{equation}
    v(E_{\gamma}, \epsilon, \theta) = \sqrt{1 - \frac{2 m_e^2 c^4}{E_{\gamma} \epsilon (1 - \cos \theta)}}
\end{equation}
and $\alpha$ is the fine-structure constant. The cross section reaches its maximum of around $1.70 \times 10^{-25}~\mathrm{cm}^2$ when $v$ is around 0.7. When $v \ll 0$ as well as $v \rightarrow 1$, the cross section reaches $0$. In other words, suppose the collision is face to face, the cross section reaches maximum when the energy of the background photon is around $(512 ~\mathrm{GeV}/E_{\gamma})~\mathrm{eV}$. Thus in case the energy of the gamma ray photon lies in the interval between 100~GeV and 10~TeV, the absorption is dominated by the interaction with the optical/infrared photons of the EBL. While if the energy of the gamma ray photon is above 100~TeV, the absorption is dominated by the interaction with CMB photons.

\begin{figure}[htb]
    \includegraphics[width = 0.5\textwidth]{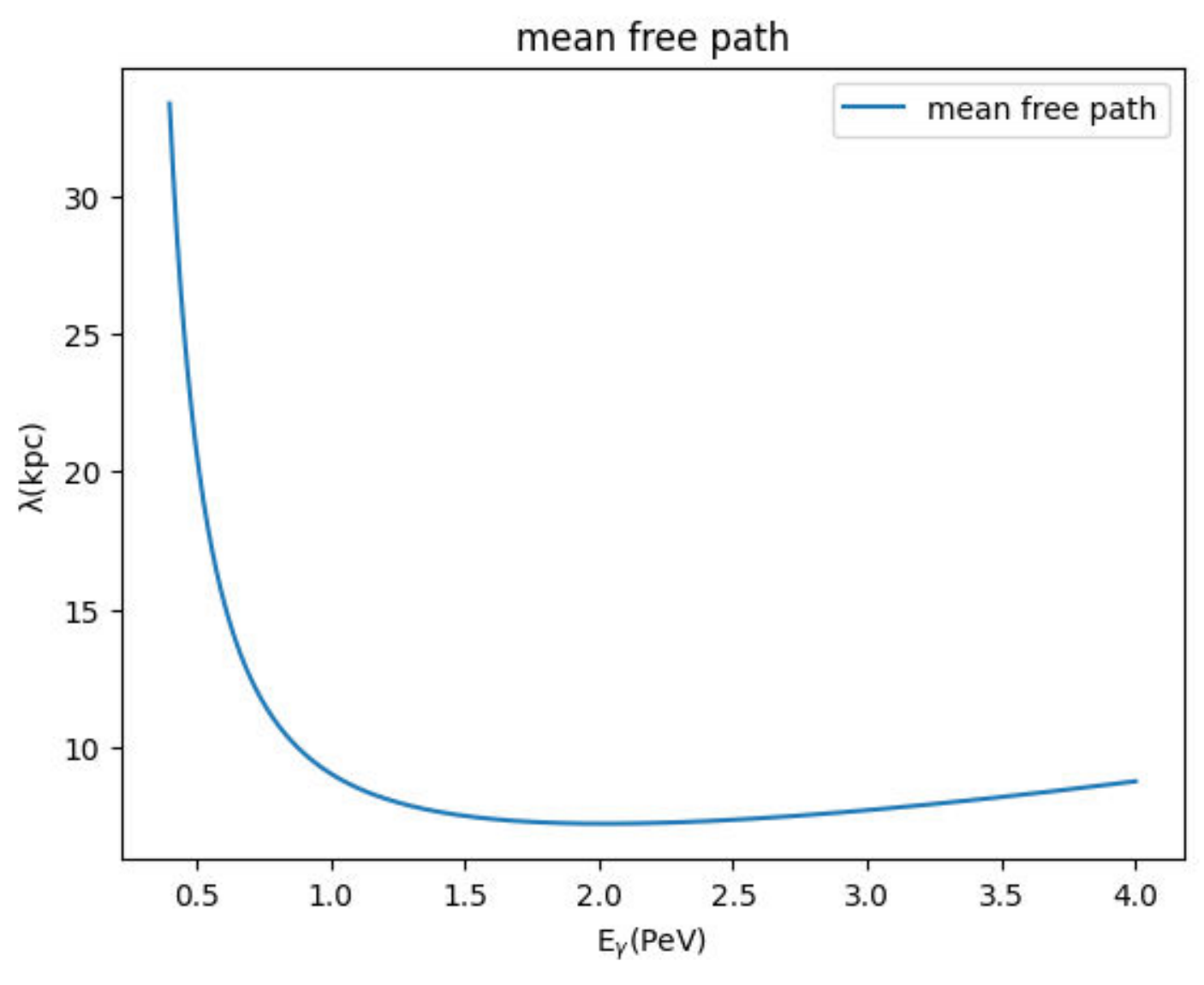}
    \caption{Mean free path of the gamma-ray photons with energy around 1~PeV due to CMB photon absorption.}
    \label{fig:mean free path}
\end{figure}

Given the gamma ray photon energy, the mean free path of the gamma ray photon due to the electron-positron pair production process in the low-energy photon background is
\begin{equation}
    \lambda^{-1}(E_{\gamma}) = \int_{-1}^{1}
    \frac{(1-\cos \theta)}{2}
    \mathrm{d} \cos \theta \int_{\epsilon_{\mathrm{th}}}^{\infty}\mathrm{d} \epsilon ~ \sigma(v(E_{\gamma}, \epsilon, \theta))  \frac{\mathrm{d} n(\epsilon)}{\mathrm{d} \epsilon},
\end{equation}
where $n(\epsilon)$ is the number density distribution of the background photons with respect to the photon energy. The CMB spectrum can be well approximated with that of black-body radiation with $T = 2.7255~\mathrm{K}$~\cite{ParticleDataGroup:2020ssz}. According to the Planck law, the photon number density distribution is thus
\begin{equation}
    \frac{\mathrm{d} n (\epsilon)}{\mathrm{d}\epsilon} = \frac{8\pi 
    }{h^3 c^3}\frac{\epsilon^2}{e^{\epsilon/(k_B T)} - 1},
\end{equation}
where $h$ is the Planck constant, $c$ is the light speed and $k_B$ is the Boltzmann constant. Fig.~\ref{fig:mean free path} shows the numerical result of the mean free path of gamma ray photons with energy around 1~PeV, which is attenuated by CMB photons. The mean free path is around 10~kpc for gamma ray photons with $E_{\gamma} = 1~\mathrm{PeV}$.

Finally, for gamma ray from a source at distance $L$ and with luminosity $L_{\mathrm{source}}$, the luminosity observed on Earth is
\begin{equation}
    L_{\mathrm{obs}} = \exp(-L/\lambda) L_{\mathrm{source}}.
\end{equation}

Recently, the Large High Altitude Air Shower Observatory (LHAASO) reported the detection of 12 UHE $\gamma$-ray sources with statistical significance $\geq 7 \sigma$~\cite{Aharonian:2021pre,Cao:2021pre1}. Among the gamma-ray-photon events detected, the event with the highest energy has an energy as high as $1.4~\mathrm{PeV}$, which is by far the most energetic gamma-ray-photon event ever been observed. Except Crab Nebulae, the origins of these PeV scale gamma-ray photon events have not yet been fully identified. Thus there still exists the possibility that some of these UHE photons come from origins that are much further than 10~kpc or even outside the Milky Way. 
The threshold anomaly of electron-positron production~\cite{Li:2021tcv,LI2021a,Li2021b} due to Lorentz invariance violation~\cite{HeMa} has been proposed
to permit the propagation of these multi-TeV-scale and PeV-scale 
photons from extragalactic origins to reach observers on the
Earth. 

The existence of axion-like particles (ALP) is suggested by several extensions of the standard model, including four-dimensional ordinary and supersymmetric models~\cite{Coriano:2006xh}, Kaluza-Klein theories~\cite{Chang:1999si} as well as superstring theories~\cite{Svrcek:2006yi, Arvanitaki:2009fg}. ALPs are very light pseudo-scalar bosons, which are usually represented with $a$. ALP is a sort of generalization of the axion, the pseudo-Goldstone boson associated with Peccei-Quinn symmetry, which is proposed as a natural solution to the strong-CP problem~\cite{Peccei:1977hh}. The key feature of ALPs is the coupling between one ALP and two photons, $a\gamma\gamma$, of which the Lagrangian is
\begin{equation}
    \mathcal{L} = \frac{1}{4} g_{a\gamma} \tilde{F}_{\mu \nu}F^{\mu \nu} ~a = g_{a\gamma} \mathbf{E}\cdot \mathbf{B} ~a.
\end{equation}
The coupling constant $g_{a\gamma}$ and the mass of the ALP $m_a$ are usually considered as independent parameters. One of the remarkable consequences of the $a\gamma \gamma$ coupling is the photon-ALP oscillation, which states that it is possible that a photon is transformed into an ALP when the former travels through transverse magnetic field and vice versa~\cite{Sikivie:1983ip, Masso:1995tw}. 

The ALP-photon oscillation scenario could help the UHE photons to travel through distance much longer than 10~kpc avoiding absorption of CMB photons through the following mechanism. Suppose the environment the UHE photons are produced is surrounded by magnetic field that is strong enough, which is the common case among typical high-energy cosmic photon factories such as pulsar, supernova remnant and active galactic nucleus (AGN), then these UHE photons may be transformed into ALPs. The resulting ALPs could travel through cosmological distance without interacting with background low-energy photons. After reaching the Milky Way the ALPs may be transformed into UHE photons again due to interaction with the Milky Way magnetic field and then be detected by observers on Earth. The scenario described here was firstly proposed to explain the excess of very-high-energy gamma ray which should undergo attenuation of the EBL photons by Ref.~\cite{DeAngelis:2007dqd} and later by Refs.~\cite{DeAngelis:2008sk, Sanchez-Conde:2009exi, Mirizzi:2009aj, Dominguez:2011xy} 
with source B-field or the intergalactic magnetic fields (IGMF), and then was developed in Ref.~\cite{referee-ref2} to include the Milky Way magnetic field.
The are also papers involving extragalactic UHE photon-ALP conversion~\cite{referee2ref2,referee2ref3,referee2ref4}, which is feasible with the LHAASO detection ability of UHE photon events.

The probability that a photon is transformed into an ALP after travelling for distance $l$ through environment with transverse magnetic field reads \cite{Sikivie:1983ip}
\begin{equation}
P_{a\gamma} =   \frac{1}{1 + (E_{\mathrm{c}}/E_{\gamma})^2} \sin^2 \Big(\frac{g_{a\gamma}B_T l}{2}\sqrt{1 +(E_{\mathrm{c}}/E_{\gamma})^2} \Big),
\label{probability}
\end{equation}
where $B_T$ is the transverse magnetic field and the critical energy $E_{\mathrm{c}}$ is defined as \cite{Mirizzi:2009aj}
\begin{equation}
\begin{aligned}
    E_{\mathrm{c}} &= \frac{m_a^2 - \omega_{\mathrm{pl}}^2}{2 g_{a\gamma}B_T}\\
    &\simeq 25\frac{|m_a^2 - \omega_{\mathrm{pl}}|}{(10^{-10} ~\mathrm{eV})^2}\Big(\frac{10^{-9}~\mathrm{Gauss}}{B_T}\Big)\Big(\frac{10^{-11} ~\mathrm{GeV}^{-1}}{g_{a\gamma}}\Big)~\mathrm{GeV},
\end{aligned}
\end{equation}
with $\omega_{\mathrm{pl}}$ being the plasma frequency of the medium 
\begin{equation}
    \omega_{\mathrm{pl}} = \sqrt{\frac{4 \pi \alpha n_e}{m_e}},
\end{equation}
here $n_e$ is the electron number density of the medium. Eq.~(\ref{probability}) shows that the maximum value of the transformation probability increases with the increasing of the energy of the photon $E_{\gamma}$. In the limit that $E \gg E_{\mathrm{c}}$, the transformation probability $P_{a \gamma}$ oscillates periodically in space with the maximum amplitude reaching one. Thus UHE photons that travel through cosmological distance provide an efficient approach for testing the existence of such scenario as well as searching for ALPs. 

The extragalactic background light (EBL) is the second,
after the CMB, most abundant part of the photon medium in
the Universe.
During the propagation of VHE and UHE photons in the Universe, their interaction with low-energy EBL photons should be also considered. 
The situation of the EBL background photon distribution could be more complicated, and as a rough estimate, we can take \(\epsilon \simeq 10^{-3}~\text{eV}\) to \(1~\text{eV}\) according to the distribution of EBL, leading to the corresponding thresholds \(E_{\text{th}}\simeq 261~\text{GeV}\) to \(261~\text{TeV}\)~\cite{lihao,Li:2022wxc}. 
A recent analysis on data suggests a photon number density $n_{\mathrm{EBL}} = 10^4~m^{-3}$ with a typical EBL photon energy of 1~eV~\cite{Brevik:2020cky}. One can take this simple EBL model to estimate the EBL influence on the VHE and UHE photon propagation in the Universe.  More explicit calculations using a realistic EBL model~\cite{lihao} suggest that cosmic photons above 261~GeV are also difficulty to reach detectors on the Earth due to their absorption by the interaction with EBL background photons if their sources are far away. 

More recently, the newly observed gamma ray burst GRB 221009A on 9 October 2022 has received extensive attention due the brightest and near-distance features of this special GRB~\cite{Zhu2022}. More strikingly, LHAASO reported~\cite{gcn32677} 
the observation of a large number of very-high energy photon events with energies larger than 500~GeV including also VHE photons with energy up to 18~TeV to be associated with this GRB 22100A. Immediately afterward it is first reported in Ref.~\cite{Li:2022wxc} that the observation of such high energy photons by LHAASO is an extraordinary result since extragalactic background light could absorb these photons severely and the flux is too weak to be observed. Then it is proposed in Ref.~\cite{Li:2022wxc} that Lorentz invariance violation induced threshold anomaly of electron-positron production~\cite{Li:2021tcv,LI2021a,Li2021b}, i.e., new physics beyond special relativity, 
can serve as an explanation of this observation. There have been also a number of proposals to explain the high energy features of GRB 221009A by axion interpretations~\cite{Galanti:2022pbg,Baktash:2022gnf,space,Nakagawa:2022wwm}. 
The suggestion of adopting the axion-photon conversion as a mechanism to explain the high energy features of GRB 221009A relies on the astrophysical environments with pronounced magnetic fields in the line of sight of the following characteristics: (i) photon-to-axion conversion upon leaving the host galaxy or cluster; (ii) non-significant re-conversion along the intergalactic medium (IG) 
in the path to observer; (iii) axion-to-photon conversion upon entry into the Milky Way. 
For (ii) to be feasible, we need to assume that the intergalactic magnetic field is weak enough. More specifically, this can be realized for~\cite{space}
\begin{equation}
\frac{g_{a\gamma}}{10^{-11}~\mathrm{GeV}^{-1}} \lesssim \left(\frac{E}{255~ \mathrm{TeV}}\right)^{-1}\left(\frac{m_a}{10^{-8}~\mathrm{eV}}\right)^{2}\left(\frac{B_{\mathrm{IG}}}{\mathrm{nGauss}}\right)^{-1}.
\end{equation}
In fact, the explicit parameter space for the coupling constant $g_{a\gamma}$ and the ALP mass $m_a$ has been constrained from various investigations~\cite{referee-ref1,referee-ref2,referee-ref3,referee-ref4,referee-ref5,ALPwebsite}, and the allowed region of the parameter space to explain the high energy photons associated with the newly observed GRB 221009A is also discussed in Ref.~\cite{space}, suggesting that the axion-photon conversion can serve as a viable mechanism to explain the high energy features of GRB 221009A. 
Here we indicate that the axion-photon conversion may also explain some of the multi-TeV  and PeV  photons already observed by LHAASO once these photons can be associated with extragalactic origins. We notice that Carpet-2 reported the coincidence of a 251~TeV photon event with GRB 221009A~\cite{Carpet-2}. This can be considered as a supplementary support to our proposal of adopting the axion-photon conversion as an alternative mechanism for the detection of extragalactic multi-TeV and PeV photons.

In summary, we discuss the detection of multi-TeV-scale and PeV-scale photons by LHAASO recently, and such multi-TeV and PeV photons may meet serious challenge due to background low-energy photon absorption if these gamma photons come from sources farther than 10~kpc or even of extragalactic origins. We also illustrate a mechanism that relies on the scenario of the oscillation between photons and the hypothetical axion-like particles (ALPs).
Due to interaction with the magnetic field surrounding the source, a considerable proportion of the VHE and UHE photons may be converted into ALPs and then travel through cosmological distance unaffected by EBL and CMB photons. Before reaching the Earth, a substantial fraction of these ALPs may be converted into VHE and UHE photons by the Galaxy magnetic field and then be detected by observers on the Earth. Therefore it is urgent to check whether the VHE and UHE photons detected by LHAASO have an extragalactic origin or not.

{\it{Acknowledgements.~}}
This work is supported by National Natural Science Foundation of China (Grant No.~12075003). We acknowledge the help discussions with Hao Li and Luohan Wang.

\end{document}